\documentclass[10pt,twocolumn]{article}

\usepackage{graphicx}
\usepackage{subcaption}
\usepackage{physics}
\usepackage{makecell}
\usepackage{diagbox}
\usepackage{xifthen}
\usepackage{url}

\newcommand{\diag}[2][]{\text{diag}_{#1}[#2]}
\newcommand{\pdiagn}[1]{\diag[n]{e^{i#1}}}
\newcommand{\proj}[1][\cdot]{\text{proj}[#1]}
\newcommand{\outphases}[2][]{D_{#1}(#2)}
\newcommand{\coutphases}[2][]{D_{#1}^\ast(#2)}
\newcommand{\outphase}[2][]{\delta_n^{#1}(#2)}

\begin{document}

\title{Fast reconstruction of programmable interferometers with intensity-only measurements}

\author{%
Bantysh B.I.$^{1}$, Chernyavskiy A.Yu.$^{1}$, Fldzhyan S.A.$^{2}$, Bogdanov Yu.I.$^{1}$ \\
\textit{\small 1 -- Kurchatov Institute National Research Centre, Moscow, Russia} \\
\textit{\small 2 -- Quantum Technology Centre, Faculty of Physics, M. V. Lomonosov Moscow State University, Moscow, Russia}\\
}

\date{bbantysh60000@gmail.com}

\maketitle

\begin{abstract}
Programmable linear optical interferometers are promising for classical and quantum applications. Their integrated design makes it possible to create more scalable and stable devices. To use them in practice, one has to reconstruct the whole device model taking the manufacturing errors into account. The inability to address individual interferometer elements complicates the reconstruction problem. A naive approach is to train the model via some complex optimization procedure. A faster optimization-free algorithm has been recently proposed [Opt. Express 31, 16729 (2023)]. However, it requires the full transfer matrix tomography while a more practical setup measures only the fields intensities at the interferometer output. In this paper, we propose the modification of the fast algorithm, which uses additional set of interferometer configurations in order to reconstruct the model in the case of intensity-only measurements. We show that it performs slightly worse than the original fast algorithm but it is more practical and still does not require intensive numerical optimization.  
\end{abstract}

\vspace{2pc}

\section{Introduction}

Linear optical interferometers are promising devices for performing quantum and classical computation \cite{minzioni2019roadmap,carolan2015universal,wang2020integrated,zhong201812,asavanant2019generation,wetzstein2020inference,zhang2021optical,deng2023gaussian}. The presence of programmable elements in them allows one to implement an arbitrary unitary transformation of the input fields. The transition to the integrated design could make such devices more stable and scalable. The light propagates in them through the waveguides and transforms via a set of static (two- or multi-port beamsplitters) and controllable (phase shifters) elements. There are many ways to arrange these elements resulting in different device size and errors robustness \cite{reck1994experimental,clements2016optimal,fldzhyan2020optimal,saygin2020robust}.

Manufacturing errors can be compensated to some extent by a proper adjustment of the controllable elements. In particular, for each required unitary matrix $V$, an optimization procedure can be performed in order to find the interferometer controls that bring it's output intensities close to the target ones \cite{hughes2018training}. Another way is to first obtain a complete model of the programmable interferometer as a function of control parameters. This allows one to subsequently perform control optimization on a computer, without additional measurements \cite{pai2019matrix,taguchi2023iterative}. The direct approach to reconstruction of the programmable interferometer model face the computationally intensive numerical optimization \cite{kuzmin2021architecture}. Recently, an optimization-free algorithm has been proposed \cite{bantysh2023fast}. It transforms the initial problem to the system of matrix equations, which could be solved efficiently by eigendecomposition. We describe this method in Section~\ref{sect:fast}.

The method requires the full tomography of the interferometer transfer matrix $V$ up to a global phase. It can be done by homodyne detection of each matrix element \cite{jacob2018direct}. This procedure, however, may be unstable towards changing the input and output ports of the interferometer. In this regard, the partial tomography up to input and output phases is more common \cite{rahimi2013direct,katamadze2021linear}. In other words, the matrices $V$ and $D_{out}VD_{in}$ ($D_{in}$ and $D_{out}$ are arbitrary diagonal unitary matrices) are treated the same. If the input phases are well controlled, one can perform  the tomography up to output phases only so the matrices $V$ and $D_{out}V$ are treated the same \cite{heilmann2015novel}. The ambiguity arises due to the intensity measurements at the interferometer output without any phase sensitivity.

Despite the fact that the model with unknown output phases has fewer parameters, the direct substitution of the measured transfer matrices into the fast reconstruction algorithm \cite{bantysh2023fast} gives incorrect results. In this paper, we propose its modification. The new algorithm requires measuring twice as many interferometer configurations, but allows, in turn, to determine the transfer matrix output phases up to a certain reference matrix. As a result, one can again reduce the model reconstruction problem to solving the system of algebraic equations via eigendecomposition. We describe the proposed algorithm in Section~\ref{sect:algorithm}.

Since the interferometer model is reconstructed up to the output phases, in Section~\ref{sect:benchmarking} we slightly modify the algorithm performance benchmarks considered in \cite{bantysh2023fast}. As we will show in Section~\ref{sect:simulation}, the proposed algorithm performs slightly worse than the original one due to additional measurements. However, it can be implemented with a more fast and stable tomography setup and thus has a lot more practical applications.

\section{Fast reconstruction algorithm}\label{sect:fast}

Let us consider the interferometer architecture by Saygin et al. \cite{saygin2020robust} containing alternaiting static $N$-port mixing layers $U_k$ and controllable $N$-port phase layers $\Phi_k$:
\begin{equation}
    V = \Phi_{K+1}U_K\Phi_K\dots\Phi_2U_1\Phi_1.
\end{equation}
If the mixing layers sufficiently mix all $N$ modes, then with $K = N$ and the absence of linear losses, this architecture is capable of providing an arbitrary unitary transformation of the input fields. Note that most universal architectures can be reduced to this one \cite{bell2021further}.

Let us briefly describe the mixing layers reconstruction procedure proposed in \cite{bantysh2023fast}. Consider the cumulative transfer matrices $C_m=U_K\dots U_m$ ($m=1,\dots,K$). We assume that there are no absolute linear losses anywhere inside the interferometer, and thus all the cumulative matrices are invertable. The matrix $C_1$ can be measured directly by setting all interferometer phases to zero and performing the tomography. To obtain $C_m$ for $m>1$, set the phases $\Phi=\pdiagn{\varphi_n}$ on the $m$-th phase layer, where $\varphi_n=\flatfrac{2\pi(n-1)}{(N+1)}$, $n=1,\dots,N$. The phases of the remaining phase layers are set to zero. The interferometer tomography for this configuration gives the corresponding transfer matrix $V_m=U_K\dots U_{m+1}U_m\Phi U_{k-1}\dots U_1$. Simple calculations can verify that
\begin{equation}\label{eq:fast_equation}
    V_m C_1^{-1} = C_m \Phi C_m^{-1}.
\end{equation}
The equation can be solved with respect to the matrix $C_m$ up to the transformation $C_m\rightarrow C_m\Lambda_m$, where $\Lambda_m$ is any invertible diagonal matrix (this transformation does not affect the measurement results in any way and cannot be detected). The columns of $C_m$ are formed by the eigenvectors of the matrix $V_m C_1^{-1}$. For their correct sorting, the phases of the corresponding eigenvalues should be compared with $\varphi_n$. The calibration errors of control phases can lead to sorting failures. In this case, non-zero phases are set only to $M$ fixed modes, which makes it possible to precisely reconstruct the corresponding $M$ columns of the matrix $C_m$. In this way, one can sequentially obtain all the columns of the matrix in the cost of increasing number of interferometer configurations to measure.

The resulting cumulative transfer matrices are then used to determine the matrices of the mixing layers:
\begin{equation}\label{eq:mixing_from_cumulative}
    U_{m<K}=C_{m+1}^{-1}C_m, \quad U_K=C_K.
\end{equation}
From these equations it is clear that the matrix of each mixing layer is determined by no more than two cumulative transfer matrices, while determining a single cumulative matrix requires measuring two transfer matrices of the entire interferometer. This explains the absence of error accumulation effect observed in \cite{bantysh2023fast}.

The fast algorithm is not inferior in accuracy to the optimization algorithm from \cite{kuzmin2021architecture}, while allowing reconstruction to be performed significantly faster. To implement it, however, it is necessary to perform the full tomography of the interferometer transfer matrix up to the global phase. This can be accomplished by homodyne detection of each matrix element \cite{jacob2018direct}. The complexity of this procedure is $O(N^2)$ since it does not allow parallel data accumulation from distinct output ports. Moreover, it can be unstable due to the need to constantly change the input and output ports.

\section{Intensity-only measurements}\label{sect:algorithm}

For a number of applications, only the field intensities at the output of the interferometer are of interest. The corresponding tomography algorithm makes it possible to determine the transfer matrix $V$ of the interferometer up to the output phases in $O(N)$ steps \cite{heilmann2015novel}. In practice the output phases of $V$ are often chosen in the way that turns the phases of the first column to zero, i.e. the tomography procedure outputs the matrix $\tilde{V}=\outphases{V}V$, where $\outphases{V}=\pdiagn{\outphase{V}}$ and $\outphase{V}=-\arg(V_{n1})$. Substituting the corresponding matrices in \eqref{eq:fast_equation} gives
\begin{equation}\label{eq:new_equation_bad}
    \tilde{V}_m \tilde{C}_1^{-1} = \outphases{V_m}C_m\Phi C_m^{-1} \coutphases{C_1},
\end{equation}
where not only the matrix $C_m$ is unknown, but also the phase matrices $\outphases{V_m}$ and $\outphases{C_1}$. This equation can no longer be solved by simple diagonalization.

Consider the cumulative transfer matrix $C_1=U_K\dots U_1$. Its tomography based on intensity measurements outputs $\tilde{C}_1=\outphases{C_1}U_K\dots U_1$. Since we aim to obtain the interferometer model up to the output phases, we can assign the phase matrix $\outphases{C_1}$ to the last mixing layer: $U_K\rightarrow \tilde{U}_K=\outphases{C_1} U_K$. Let us denote the new cumulative transfer matrices as $\tilde{C}_m=\tilde{U}_K\dots U_m$.

As before, to obtain the matrix $\tilde{C}_m$ for $m>1$, we perform the tomography of the interferometer, in which the phases $\Phi$ are set on the $m$-th phase layer. Let us rewrite the equation \eqref{eq:new_equation_bad} in the following form:
\begin{equation}\label{eq:x_eq}
    X = \tilde{V}_m \tilde{C}_1^{-1} = \outphases[C_1]{V_m} \tilde{C}_m \Phi \tilde{C}_m^{-1},
\end{equation}
where $\outphases[C_1]{V}=\outphases{V}\coutphases{C_1}=\diag[n]{\exp(i\delta_n^{C_1}[V])}$, and $\delta_n^{C_1}[V]=\delta_n[V]-\delta_n[C_1]$ are the output phase shifts relative to the matrix $C_1$. Next, consider the configuration, in which conjugate phases $\Phi^\ast$ are set on the $m$-th phase layer: $V_m^c=U_K\dots U_{m+1}U_m\Phi^\ast U_{k-1}\dots U_1$. The tomography of the interferometer gives the matrix $\tilde{V}_m^c = \outphases{V_m^c}V_m^c$. Let us construct the following equation from $\tilde{C}_1$ and $\tilde{V}_m^c$:
\begin{equation}\label{eq:y_eq}
    Y=\tilde{C}_1(\tilde{V}_m^c)^{-1} = \tilde{C}_m \Phi \tilde{C}_m^{-1} \coutphases[C_1]{V_m^c}.
\end{equation}
Comparing \eqref{eq:x_eq} and \eqref{eq:y_eq} one obtains
\begin{equation}
    X=\outphases[C_1]{V_m} Y \outphases[C_1]{V_m^c}.
\end{equation}
Evaluating the phases of $X$ gives
\begin{equation}\label{eq:xy_args}
    \arg(x_{pq}) = \delta_p^{C_1}[V_m] + \arg(y_{pq}) + \delta_q^{C_1}[V_m^c].
\end{equation}
From these equations one can separately determine $\outphases[C_1]{V_m}$ up to a global phase and solve \eqref{eq:x_eq} with respect to $\tilde{C}_m$ via diagonalization. Another solution could be found by determining $\outphases[C_1]{V_m^c}$ from \eqref{eq:xy_args} and solving \eqref{eq:y_eq}. We combine two solutions by taking their matrix geometric mean (e.g. see \cite{iannazzo2016geometric}). Next, by determining $\tilde{C}_m$ for all $m$, one can calculate the transfer matrices of all mixing layers similarly to \eqref{eq:mixing_from_cumulative}. The last mixing layer $\tilde{U}_K$ is defined up to the output phases, which are undetectable in the case intensity-only measurements.

To reduce the role of phase layers calibration one can also employ the strategy of partial reconstruction: the number of non-zero phases in $\Phi$ and $\Phi^\ast$ is limited by $M$. Then only the $M$ columns of the matrix $\tilde{C}_m$ are reconstructed at once.

Finally, considering the case of homogeneous losses one projects the resulting transfer matrices of mixing layers onto the set of unitary matrices.

\begin{figure}[!ht]
\centering
\includegraphics[width=0.8\linewidth]{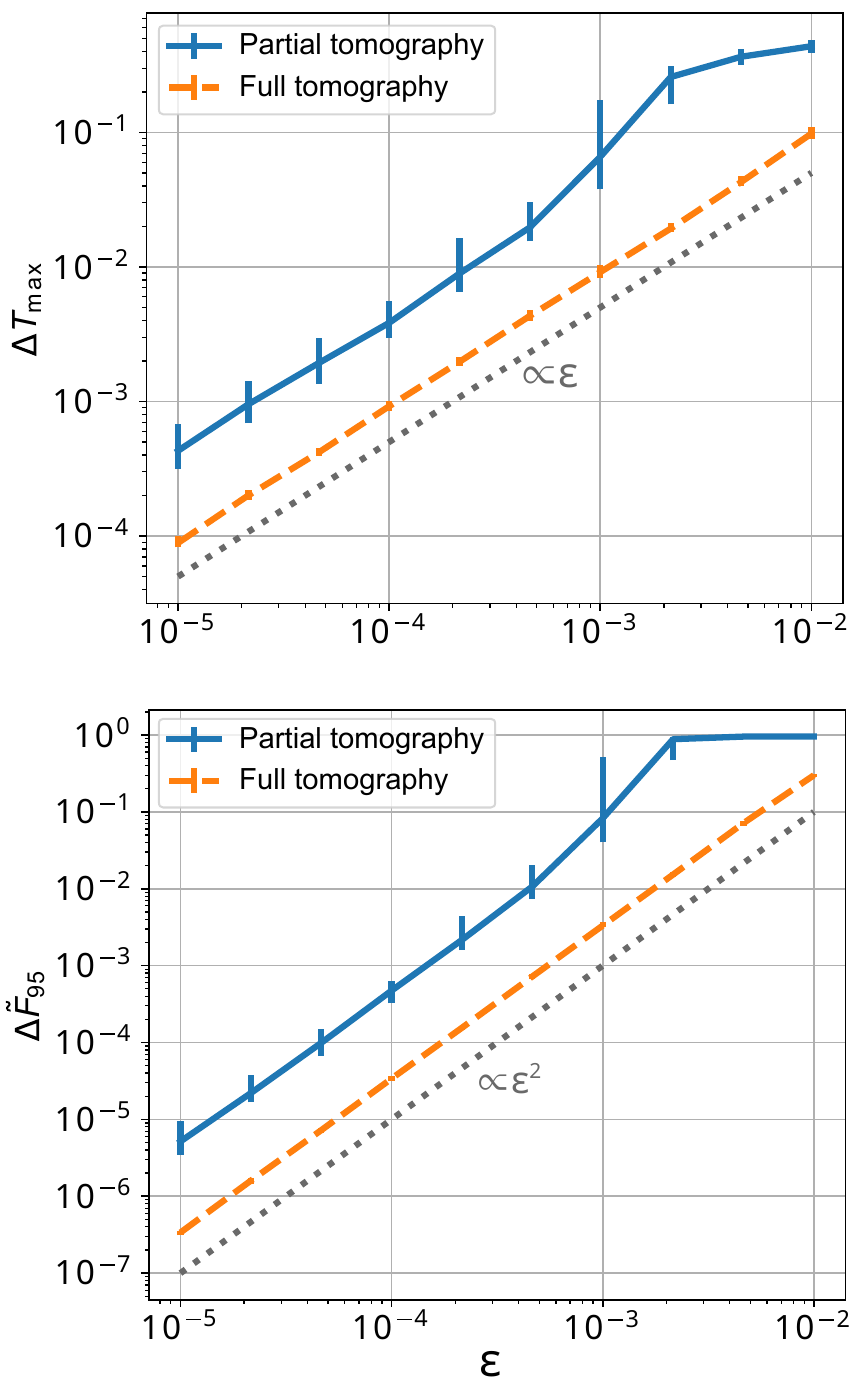}
\caption{The dependence of the maximal error of transmission coefficients estimation (top) and the error of the transfer matrix prediction (bottom) on the tomography error. The number of input and output modes is $N = 10$, the number of simultaneously determined columns of the cumulative matrices is $M = N$. The median values for 100 independent numerical experiments are shown. Confidence intervals show the lower and upper quartiles. The deviations for the full tomography case almost vanish.}
\label{fig:err_vs_noise}
\end{figure}

\section{Benchmarking methodology}\label{sect:benchmarking}

We now turn to analyzing the performance of the proposed algorithm using the benchmarks from \cite{bantysh2023fast}. Let us introduce the parameter $\gamma$ for the error of mixing layers manufacture:
\begin{equation}
    U_k \rightarrow \proj[U_H + \gamma \mathcal{G}],
\end{equation}
where $U_H$ is $N$-dimensional Hadamard transform, $\mathcal{G}$ is the complex Ginibre ensemble ($N\times N$ complex-valued matrix of independent standard Gaussian random variables \cite{lacroix2019intermediate}), and $\proj$ is the projection onto the set of unitary matrices. Below we consider $\gamma = 0.1$. We simulate the noise of tomography of each transfer matrix $V$ as
\begin{equation}\label{eq:noise_tomo}
    V \rightarrow \proj[V + \varepsilon \sqrt{L} \mathcal{G}].
\end{equation}
Here $\varepsilon$ characterizes the noise strength, and $L$ is the total number of transfer matrices that one has to measure to implement the algorithm. The correction $\sqrt{L}$ is used to compare different measurement protocols under conditions of a fixed measurement time. The protocol proposed in this work requires the tomography of $2L_F-1$ interferometer configurations, where $L_F$ is the number of configurations for the fast algorithm \cite{bantysh2023fast}. Therefore, the tomography error of each transfer matrix is approximately $\sqrt2$ times greater for the proposed algorithm compared to the fast one. Simulating the tomography based on intensity-only measurements we perform the transformation $V\rightarrow \tilde{V}=\outphases{V}V$ in addition to \eqref{eq:noise_tomo}.

We consider three different benchmarks. The first one is the maximal error of transmission coefficients estimation over the all mixing layers:
\begin{equation}\label{eq:benchmark_dt}
    \Delta T_{\max} = \max_{\substack{k=1,\dots,K\\m,n=1,\dots,N}}{\abs{[T_k^{\text{true}}]_{mn} - [T_k^{\text{rec}}]_{mn}}},
\end{equation}
where $[T_k]_{mn} = \abs{[U_k]_{mn}}^2$ is the transmission coefficient between the $n$-th input mode and the $m$-th output mode in the $k$-th mixing layer. The superscripts \textit{true} and \textit{rec} correspond to the true and reconstructed matrices respectively.

We also consider the fidelity of transfer matrix prediction:
\begin{equation}\label{eq:benchmark_df}
    F = \frac{\abs{\Tr(V_{\text{true}}^\dagger V_{\text{pred}})}^2}{\Tr(V_{\text{true}}^\dagger V_{\text{true}})\Tr(V_{\text{pred}}^\dagger V_{\text{pred}})},
\end{equation}
where $V_{\text{true}}$ is the true transfer matrix for a given set of control parameters, and $V_{\text{pred}}$ is the matrix predicted by the obtained interferometer model. Since we aim to reconstruct the model up to the output phases we would like to maximize the fidelity between $\tilde{V}_{\text{true}}=D_1 V_{\text{true}}$ and $\tilde{V}_{\text{pred}}=D_1 V_{\text{pred}}$ over diagonal unitary matrices $D_1$ and $D_2$. The denominator of \eqref{eq:benchmark_df} does not depend on $D_1$ and $D_2$ so one has to maximize the nominator. This can be done by the wavefront matching method \cite{sakamaki2007new,kupianskyi2023high}. In particular, $|\text{Tr}(V_{\text{true}}^\dagger D_1^\ast D_2 V_{\text{pred}})|=|\Tr(WD)|$, where $W=V_{\text{pred}} V_{\text{true}}^\dagger$ and $D=D_1^\ast D_2$. The maximum of $|\Tr(WD)|=|\sum_{n}{W_{nn}D_{nn}}|$ is achieved when the diagonal elements of $W$ sum up with the same phase, so $\max_D|\Tr(WD)|=\sum_{n}{|W_{nn}|}$ and
\begin{equation}\label{eq:benchmark_df_out}
    \tilde{F} = \frac{\qty(\sum_{n}{\abs{\sum_{m}{V_{\text{true},nm}^\ast V_{\text{pred},nm}}}})^2}{\Tr(V_{\text{true}}^\dagger V_{\text{true}})\Tr(V_{\text{pred}}^\dagger V_{\text{pred}})},
\end{equation}
Thus, we consider the benchmark value $\Delta\tilde{F}_{95}$, which is the 95th percentile of $\Delta\tilde{F} = 1-\tilde{F}$, calculated from 1000 interferometer configurations, where each phase of each phase layer is selected randomly and uniformly in the interval $[0,2\pi)$.

Finally, we estimate the computation time of the algorithm. The whole simulation was done using \texttt{ILOptics} software library \cite{lib_github}.

\begin{figure}[ht]
\centering
\includegraphics[width=0.8\linewidth]{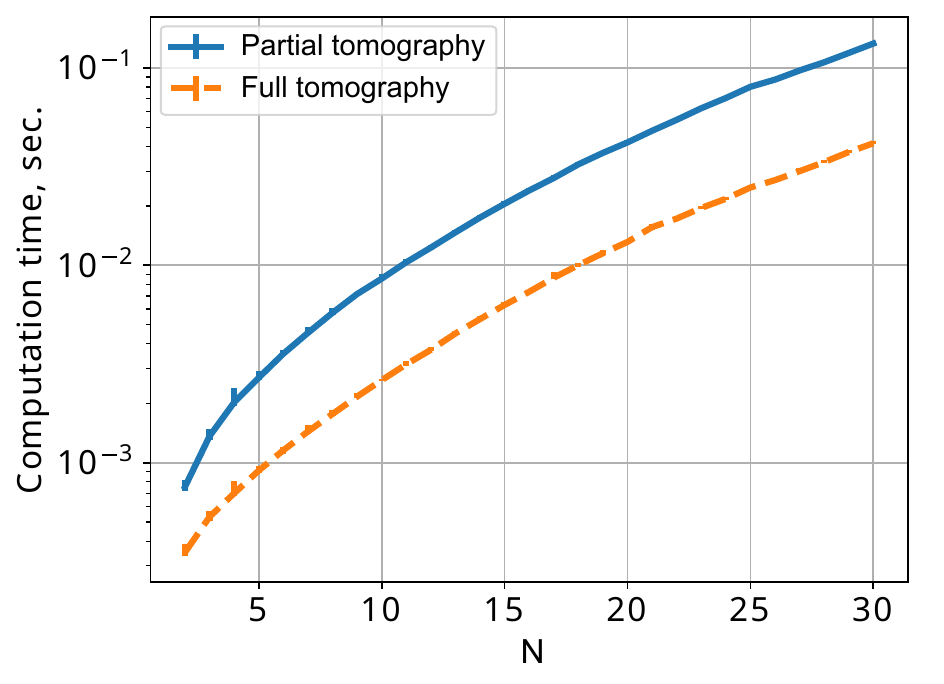}
\caption{Computation time for the interferometer model reconstruction versus the number of input and output modes. The number of simultaneously determined columns of the cumulative matrix is $M=N$. The median values for 100 independent numerical experiments are shown. Confidence intervals show the lower and upper quartiles. The computation was done on \texttt{Intel(R) Core(TM) i5-7200U CPU @ 2.50GHz}.}
\label{fig:time_vs_dim}
\end{figure}

\section{Simulation results}\label{sect:simulation}

Let us study the performance of the fast algorithm based on two types of transfer matrix tomography:
\begin{itemize}
    \item the partial tomography based on intensity-only measurements (output phases are unknown);
    \item the full tomography (considered in \cite{bantysh2023fast}).
\end{itemize}
We consider the case of homogeneous losses, so the reconstructed cumulative transfer matrices are projected onto the set of unitary matrices.

Figure~\ref{fig:err_vs_noise} shows the results of numerical simulation for the case $N = 10$. One can observe the dependencies $\Delta T_{\max}\propto\varepsilon$ and $\Delta\tilde{F}_{95}\propto\varepsilon^2$ both in the cases of partial and full tomography. At the same time, the proportionality coefficients are different: in order to obtain the same benchmark values, the partial tomography error must be approximately half an order of magnitude smaller. Moreover, the algorithm based on partial tomography has a higher spread of benchmark values. This can be explained by two factors. First, intensity-only measurements limits the amount of information that can be obtained from the tomography of a single transfer matrix. Second, the reconstruction of a single cumulative matrix requires the tomography of three interferometer configurations, instead of two, compared with algorithm bases on the full tomography: one additional configuration is associated with conjugate phases $\Phi^\ast$. Since the tomography of each matrix is accompanied by fluctuations, this makes an additional contribution to the model reconstruction inaccuracy.

Note that the above calculations do not take into account the time spent on the tomography procedure itself. The parallel measurement of intensities proposed in \cite{heilmann2015novel} allows the tomography to be performed $N$ times faster compared to the homodyne based method described in \cite{jacob2018direct}. This may also affect the magnitude of tomography result fluctuations under conditions of a fixed measurement time.

The proposed algorithm requires more operations than the original algorithm from \cite{bantysh2023fast}, which slightly increases the total model reconstruction time (Figure~\ref{fig:time_vs_dim}). However, it still does not use numerical optimization and takes just about $0.1$ seconds on \texttt{Intel(R) Core(TM) i5-7200U CPU @ 2.50GHz} in the case $N=K=30$. To compare, for $K=N$ the unitary interferometer model contains $N^2(N-1)=26,100$ independent real parameters to optimize.

\section{Conclusion}

In this work, we proposed an algorithm for reconstructing the model of a programmable linear optical interferometer. Compared with the previously published fast algorithm \cite{bantysh2023fast}, we made it possible to estimate the interferometer mixing layers even in the case of partial tomography, when there is no information about the fields output phases. Using a set of benchmarks, we showed that the new algorithm performs qualitatively the same as the previous one. However, since it requires more interferometer configurations to measure, the benchmark values have slightly worse values. Nevertheless, the corresponding experimental procedure is faster and more suitable for practical applications.

\section{Acknowledgments}

This work was supported by the State Program no. FFNN-2022-0016. S~.A.~Fldzhyan is grateful to the BASIS foundation (Project No. 23-2-10-15-1).

\bibliographystyle{unsrt}
\bibliography{article.bib}

\begin{thebibliography}{10}

\bibitem{minzioni2019roadmap}
Paolo Minzioni, Cosimo Lacava, Takasumi Tanabe, Jianji Dong, Xiaoyong Hu, Gyorgy Csaba, Wolfgang Porod, Ghanshyam Singh, Alan~E Willner, Ahmed Almaiman, et~al.
\newblock Roadmap on all-optical processing.
\newblock {\em Journal of Optics}, 21(6):063001, 2019.

\bibitem{carolan2015universal}
Jacques Carolan, Christopher Harrold, Chris Sparrow, Enrique Mart{\'\i}n-L{\'o}pez, Nicholas~J Russell, Joshua~W Silverstone, Peter~J Shadbolt, Nobuyuki Matsuda, Manabu Oguma, Mikitaka Itoh, et~al.
\newblock Universal linear optics.
\newblock {\em Science}, 349(6249):711--716, 2015.

\bibitem{wang2020integrated}
Jianwei Wang, Fabio Sciarrino, Anthony Laing, and Mark~G Thompson.
\newblock Integrated photonic quantum technologies.
\newblock {\em Nature Photonics}, 14(5):273--284, 2020.

\bibitem{zhong201812}
Han-Sen Zhong, Yuan Li, Wei Li, Li-Chao Peng, Zu-En Su, Yi~Hu, Yu-Ming He, Xing Ding, Weijun Zhang, Hao Li, et~al.
\newblock 12-photon entanglement and scalable scattershot boson sampling with optimal entangled-photon pairs from parametric down-conversion.
\newblock {\em Physical review letters}, 121(25):250505, 2018.

\bibitem{asavanant2019generation}
Warit Asavanant, Yu~Shiozawa, Shota Yokoyama, Baramee Charoensombutamon, Hiroki Emura, Rafael~N Alexander, Shuntaro Takeda, Jun-ichi Yoshikawa, Nicolas~C Menicucci, Hidehiro Yonezawa, et~al.
\newblock Generation of time-domain-multiplexed two-dimensional cluster state.
\newblock {\em Science}, 366(6463):373--376, 2019.

\bibitem{wetzstein2020inference}
Gordon Wetzstein, Aydogan Ozcan, Sylvain Gigan, Shanhui Fan, Dirk Englund, Marin Solja{\v{c}}i{\'c}, Cornelia Denz, David~AB Miller, and Demetri Psaltis.
\newblock Inference in artificial intelligence with deep optics and photonics.
\newblock {\em Nature}, 588(7836):39--47, 2020.

\bibitem{zhang2021optical}
Hui Zhang, Mile Gu, XD~Jiang, Jayne Thompson, Hong Cai, S~Paesani, R~Santagati, A~Laing, Y~Zhang, MH~Yung, et~al.
\newblock An optical neural chip for implementing complex-valued neural network.
\newblock {\em Nature communications}, 12(1):457, 2021.

\bibitem{deng2023gaussian}
Yu-Hao Deng, Yi-Chao Gu, Hua-Liang Liu, Si-Qiu Gong, Hao Su, Zhi-Jiong Zhang, Hao-Yang Tang, Meng-Hao Jia, Jia-Min Xu, Ming-Cheng Chen, et~al.
\newblock Gaussian boson sampling with pseudo-photon-number resolving detectors and quantum computational advantage.
\newblock {\em arXiv preprint arXiv:2304.12240}, 2023.

\bibitem{reck1994experimental}
Michael Reck, Anton Zeilinger, Herbert~J Bernstein, and Philip Bertani.
\newblock Experimental realization of any discrete unitary operator.
\newblock {\em Physical review letters}, 73(1):58, 1994.

\bibitem{clements2016optimal}
William~R Clements, Peter~C Humphreys, Benjamin~J Metcalf, W~Steven Kolthammer, and Ian~A Walmsley.
\newblock Optimal design for universal multiport interferometers.
\newblock {\em Optica}, 3(12):1460--1465, 2016.

\bibitem{fldzhyan2020optimal}
Suren~A Fldzhyan, M~Yu Saygin, and Sergei~P Kulik.
\newblock Optimal design of error-tolerant reprogrammable multiport interferometers.
\newblock {\em Optics Letters}, 45(9):2632--2635, 2020.

\bibitem{saygin2020robust}
M~Yu Saygin, IV~Kondratyev, IV~Dyakonov, SA~Mironov, SS~Straupe, and SP~Kulik.
\newblock Robust architecture for programmable universal unitaries.
\newblock {\em Physical review letters}, 124(1):010501, 2020.

\bibitem{hughes2018training}
Tyler~W Hughes, Momchil Minkov, Yu~Shi, and Shanhui Fan.
\newblock Training of photonic neural networks through in situ backpropagation and gradient measurement.
\newblock {\em Optica}, 5(7):864--871, 2018.

\bibitem{pai2019matrix}
Sunil Pai, Ben Bartlett, Olav Solgaard, and David~AB Miller.
\newblock Matrix optimization on universal unitary photonic devices.
\newblock {\em Physical review applied}, 11(6):064044, 2019.

\bibitem{taguchi2023iterative}
Yoshitaka Taguchi, Yunzhuo Wang, Ryota Tanomura, Takuo Tanemura, and Yasuyuki Ozeki.
\newblock Iterative configuration of programmable unitary converter based on few-layer redundant multiplane light conversion.
\newblock {\em Physical Review Applied}, 19(5):054002, 2023.

\bibitem{kuzmin2021architecture}
Sergei Kuzmin, Ivan Dyakonov, and Sergei Kulik.
\newblock Architecture agnostic algorithm for reconfigurable optical interferometer programming.
\newblock {\em Optics Express}, 29(23):38429--38440, 2021.

\bibitem{bantysh2023fast}
Boris Bantysh, Konstantin Katamadze, Andrey Chernyavskiy, and Yurii Bogdanov.
\newblock Fast reconstruction of programmable integrated interferometers.
\newblock {\em Optics Express}, 31(10):16729--16742, 2023.

\bibitem{jacob2018direct}
Kevin~Valson Jacob, Anthony~E Mirasola, Sushovit Adhikari, and Jonathan~P Dowling.
\newblock Direct characterization of linear and quadratically nonlinear optical systems.
\newblock {\em Physical Review A}, 98(5):052327, 2018.

\bibitem{rahimi2013direct}
Saleh Rahimi-Keshari, Matthew~A Broome, Robert Fickler, Alessandro Fedrizzi, Timothy~C Ralph, and Andrew~G White.
\newblock Direct characterization of linear-optical networks.
\newblock {\em Optics express}, 21(11):13450--13458, 2013.

\bibitem{katamadze2021linear}
KG~Katamadze, GV~Avosopiants, AV~Romanova, Yu~I Bogdanov, and SP~Kulik.
\newblock Linear optical circuits characterization by means of thermal field correlation measurement.
\newblock {\em Laser Physics Letters}, 18(7):075201, 2021.

\bibitem{heilmann2015novel}
Ren{\'e} Heilmann, Markus Gr{\"a}fe, Stefan Nolte, and Alexander Szameit.
\newblock A novel integrated quantum circuit for high-order w-state generation and its highly precise characterization.
\newblock {\em Science bulletin}, 60:96--100, 2015.

\bibitem{bell2021further}
B.~A. Bell and I.~A. Walmsley.
\newblock Further compactifying linear optical unitaries.
\newblock {\em APL Photonics}, 6(7):070804, 2021.

\bibitem{iannazzo2016geometric}
Bruno Iannazzo.
\newblock The geometric mean of two matrices from a computational viewpoint.
\newblock {\em Numerical Linear Algebra with Applications}, 23(2):208--229, 2016.

\bibitem{lacroix2019intermediate}
Bertrand Lacroix-A-Chez-Toine, Jeyson Andr{\'e}s~Monroy Garz{\'o}n, Christopher Sebastian~Hidalgo Calva, Isaac~P{\'e}rez Castillo, Anupam Kundu, Satya~N Majumdar, and Gr{\'e}gory Schehr.
\newblock Intermediate deviation regime for the full eigenvalue statistics in the complex ginibre ensemble.
\newblock {\em Physical Review E}, 100(1):012137, 2019.

\bibitem{sakamaki2007new}
Yohei Sakamaki, Takashi Saida, Toshikazu Hashimoto, and Hiroshi Takahashi.
\newblock New optical waveguide design based on wavefront matching method.
\newblock {\em Journal of lightwave technology}, 25(11):3511--3518, 2007.

\bibitem{kupianskyi2023high}
Hlib Kupianskyi, Simon~AR Horsley, and David~B Phillips.
\newblock High-dimensional spatial mode sorting and optical circuit design using multi-plane light conversion.
\newblock {\em APL Photonics}, 8(2), 2023.

\bibitem{lib_github}
B.~Bantysh.
\newblock The package for simulating and training integrated linear optical devices.
\newblock \url{https://github.com/PQCLab/ILOptics}.

\end{thebibliography}

\end{document}